\documentclass[aps,twocolumn,pra,showpacs,tightenlines]{revtex4-1}
\usepackage{amsmath}
\usepackage{graphicx}
\usepackage{color}
\usepackage{amsfonts}
\usepackage{txfonts}
\usepackage[colorlinks,citecolor=blue]{hyperref}

\begin{document}

\title{Two-photon scattering in mixed cavity optomechanics}
\author{Dong-Cheng Chen}
\affiliation{Key Laboratory of Low-Dimensional Quantum Structures and Quantum Control of
Ministry of Education, Key Laboratory for Matter Microstructure and Function
of Hunan Province, Department of Physics and Synergetic Innovation Center
for Quantum Effects and Applications, Hunan Normal University, Changsha
410081, China}
\author{Yue-Hui Zhou}
\affiliation{Key Laboratory of Low-Dimensional Quantum Structures and Quantum Control of
Ministry of Education, Key Laboratory for Matter Microstructure and Function
of Hunan Province, Department of Physics and Synergetic Innovation Center
for Quantum Effects and Applications, Hunan Normal University, Changsha
410081, China}
\author{Jin-Feng Huang}
\email{Corresponding author: jfhuang@hunnu.edu.cn}
\affiliation{Key Laboratory of Low-Dimensional Quantum Structures and Quantum Control of
Ministry of Education, Key Laboratory for Matter Microstructure and Function
of Hunan Province, Department of Physics and Synergetic Innovation Center
for Quantum Effects and Applications, Hunan Normal University, Changsha
410081, China}
\author{Jie-Qiao Liao}
\email{Corresponding author: jqliao@hunnu.edu.cn}
\affiliation{Key Laboratory of Low-Dimensional Quantum Structures and Quantum Control of
Ministry of Education, Key Laboratory for Matter Microstructure and Function
of Hunan Province, Department of Physics and Synergetic Innovation Center
for Quantum Effects and Applications, Hunan Normal University, Changsha
410081, China}
\date{\today }

\begin{abstract}
We study two-photon scattering in a mixed cavity optomechanical system, which is composed of a single-mode cavity field coupled to a single-mode mechanical oscillation via both the first-order and quadratic optomechanical interactions. By solving the scattering problem within the Wigner-Weisskopf framework, we obtain the analytical scattering state and find four physical processes associated with the two-photon scattering in this system. We calculate the two-photon scattering spectrum and find that two-photon frequency anticorrelation can be induced in the scattering process. We also establish the relationship between the parameters of the mixed cavity optomechanical system and the characteristics of the two-photon scattering spectrum. This work not only provides a scattering means to create correlated photon pairs, but also presents a spectrometric method to characterize the optomechanical systems.
\end{abstract}

\date{\today }
\maketitle

\global\columnwidth20.5pc \global\hsize\columnwidth\global\linewidth%
\columnwidth
\global\displaywidth\columnwidth

\section{INTRODUCTION}

The photon scattering process not only provides an efficient means to characterize the physical properties of the scattering target, but also opens a new route to manipulate the state of the scattered photons~\cite{Sheremet2021}. Recently, the few-photon scattering physics has been studied systematically in various quantum optical systems, such as cavity quantum electrodynamical (QED) systems~\cite{Chen2004PRA} and waveguide QED systems~\cite{Shen2007PRL,ZhouL2008PRL,Fan2010PRA,Shi2009PRB,Sun2011PRA,Hurst2018PRA,Shen2007PRL,Shen2007PRA,Tsoi2009PRA,Witthaut2010,Nysteen2015NJP,Das2018PRA,Das2018,Rephaeli2011PRA,Cheng2017PRA}. Usually, nonlinear quantum systems are used to play the roles of scattering targets. These considered scattering targets include a single emitter~\cite{Shen2007PRL,ZhouL2008PRL,Fan2010PRA,Shi2009PRB,Sun2011PRA,Hurst2018PRA,Shen2007PRL,Shen2007PRA,Tsoi2009PRA,Witthaut2010,Nysteen2015NJP,Das2018PRA,Das2018} or several emitters~\cite{Rephaeli2011PRA,Cheng2017PRA}, the Kerr-type nonlinear cavity~\cite{Liao2010PRA,Xu2013PRL,Xu2014PRA}, and optomechanical cavities~\cite{Liao2012PRA,Liao2013PRA,Jia2013PRA,Liao2014Sci,Ng2016PRA,Qiao2017PRA}. Owing to the nonlinear nature of these scattering targets, photon correlation will be induced in the few-photon scattering processes~\cite{Shen2007PRL,Shi2009PRB,Sun2011PRA,Shen2007PRA,Liao2010PRA,Liao2013PRA,Kojima2003PRA,Richter2009PRL,Roy2010PRB,Roy2011PRL,Zheng2012PRA,Ke2019PRL}. These correlated photons have wide potential applications in both the demonstration of the fundamentals of quantum theory~\cite{Glauber1963PRL,Aspect1981PRL} and the development of modern quantum information processing~\cite{Nielsen2000}.

Cavity optomechanical systems~\cite{Kippenberg2008,Aspelmeyer2014,Bowen2016}, as typical nonlinear quantum optical platforms, have been widely used to manipulate the quantum properties of photons via mechanical means. It has been shown that both the photon blockade~\cite{Rabl2011,Liao2013,Xu2013PRA} and optical quadrature squeezing~\cite{Safavi2013Nature,Purdy2013PRX,Aggarwal2020}
can be created by optomechanical interactions. Meanwhile, the optomechanically induced transparency~\cite{Agarwal2010PRA,Weis2010,Safavi2011Nat} has been demonstrated in optomechanincal systems. In particular, one of us (Liao) and coworkers~\cite{Liao2013PRA} have shown that photon correlation can be induced by scattering two free photons by a first-order cavity optomechanical system. Based on the fact that the mixed cavity optomechanical system~\cite{Rocheleau2010Nat,Xuereb2013PRA,Zhang2014PRA,Hauer2018PRA,Zhang2018PRA,Brunelli2018PRA,Zhou2019,Sainadh2020OL} is a general optomechanical platform, which includes the first-order optomechanical systems~\cite{Law1995PRA} and quadratic optomechanical systems~\cite{Thompson2008,Sankey2010,Bhattacharya2008PRA,Shi2013PRA,Liao2014Sci} as two special cases, a natural generalization is to study few-photon scattering in the mixed cavity optomechanical system.

In this paper, we consider a mixed cavity optomechanical system coupled to the continuous fields outside the cavity. Two free photons in the Lorentzian wavepacket~\cite{Liao2013PRA} are injected into the cavity. By solving the dynamics of the whole system including the continuous fields and the mixed cavity optomechanical system within the Wigner-Weisskopf framework, we obtain the long-time scattering state of the photons, and then the two-photon scattering spectrum is calculated analytically. Two-photon frequency anticorrelation is confirmed in the two-photon scattering spectrum in various parameter cases. By analyzing the eigen-energy spectrum of the mixed optomechanical system, we build the connection between the system parameters and the spectral features. This work will provide a new means to create photon correlation and suggest a spectroscopic method to characterize the mixed cavity optomechanical system.

The rest of this paper is organized as follows. In Sec. II, we introduce the mixed cavity optomechanical model and present the Hamiltonians. In Sec. III, we study the two-photon dynamics of the system and derive the equations of motion for these probability amplitudes. In Sec. IV, we obtain the long-time solution of these probability amplitudes within the Wigner-Weisskopf framework. In Sec. V, we present two-photon scattering spectrum and analyze the spectral features when the system works in various parameter regimes. Finally, we conclude this work in Sec. VI.
\begin{figure}[tbph]
\center
\includegraphics[width=0.48 \textwidth]{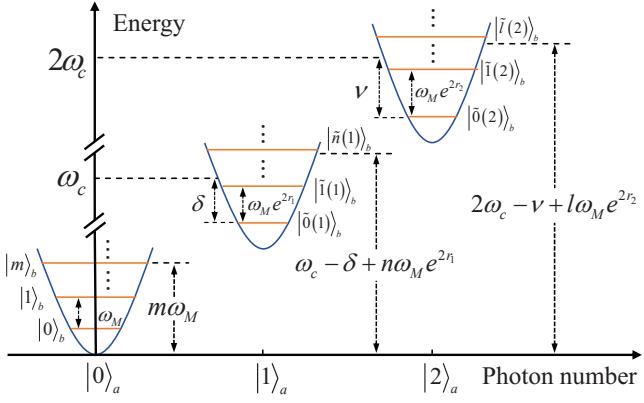}
\caption{(Color online) Diagram of the eigen-energy levels of the mixed cavity optomechanical system limited within the zero-, one-, and two-photon subspaces.}
\label{energylevel}
\end{figure}

\section{Physical system and Hamiltonian}

We consider a mixed optomechanical system, which is composed of a single-mode cavity field coupled to a single-mode mechanical oscillation via both the first-order and quadratic optomechanical interactions. The Hamiltonian of the mixed optomechanical system reads $\left( \hbar=1\right) $
\begin{equation}
H_{\text{mop}} =\omega _{c}a^{\dag }a+\omega _{M}b^{\dag }b+g_{1}a^{\dag
}a(b^{\dag }+b)+g_{2}a^{\dag }a(b^{\dag }+b)^{2},
\end{equation}%
where $a$ ($a^{\dagger }$) and $b$ ($b^{\dagger }$) are, respectively, the annihilation (creation) operators of the cavity field and the mechanical mode, with the corresponding resonance frequencies $\omega _{c}$ and $\omega_{M}$. The parameters $g_{1}$ and $g_{2}$ are the coupling strengths of the first-order and quadratic optomechanical interactions associated with a single photon between the cavity field and the mechanical oscillation, respectively.

Let us denote $|m\rangle _{a}$ ($m=0,1,2,...$) and $|j\rangle _{b}$ ($j=0,1,2,...$) as the number states of the cavity field and the mechanical oscillation, respectively. Then the eigen-equation of the Hamiltonian $H_{\text{mop}}$ can be derived as~\cite{Zhou2019}
\begin{equation}
H_{\text{mop}}|m\rangle _{a}|\tilde{j}\left( m\right) \rangle
_{b}=E_{m,j}|m\rangle _{a}|\tilde{j}\left( m\right) \rangle _{b},
\label{eigen-equation}
\end{equation}
where the eigenvalues are given by
\begin{equation}
E_{m,j}=m\omega _{c}-\frac{g_{1}^{2}e^{-4r_{m}}}{\omega _{M}}m^{2}+j\omega
_{M}e^{2r_{m}}+\frac{\omega _{M}}{2}(e^{2r_{m}}-1).
\label{eigen-value}
\end{equation}%
The states $|\tilde{j}\left( m\right) \rangle _{b}$ in Eq.~(\ref{eigen-equation}) are the $m$-photon squeezed and displaced number states of the mechanical mode, and these states are defined by%
\begin{equation}
|\tilde{j}\left( m\right) \rangle _{b}=S(r_{m})D(\alpha _{m})\left\vert
j\right\rangle _{b},
\end{equation}%
where $S(r_{m})=\exp [r_{m}(b^{2}-b^{\dag 2})/2]$ is a squeezing operator with the squeezing factor $r_{m}=\ln (4g_{2}m/\omega_{M}+1)/4$, and $D(\alpha _{m})=\exp [\alpha _{m}(b^{\dag }-b)]$ is a displacement operator with the displacement amplitude $\alpha _{m}=-g_{1}e^{-3r_{m}}m/\omega _{M}$. In particular, when there is no photon in the cavity, we have $|\tilde{j}\left( 0\right) \rangle _{b}=\left\vert j\right\rangle _{b}$. It can be seen from Eq.~(\ref{eigen-value}) that when the cavity field contains different numbers of photons, the mechanical degree of freedom has different energy-level structures. For below convenience, in Fig.~\ref{energylevel} we show the energy-level structure of the mixed optomechanical cavity limited in the zero-, one-, and two-photon subspaces.

To describe the two-photon scattering state, the continuous fields outside the cavity should be included in the system. To this end, we model the fields outside the cavity as a bosonic bath, and the couplings between the cavity field and these continuous fields are described by the photon-hopping interaction. The Hamiltonian of the whole system including the optomechanical cavity and these continuous fields can be written as
\begin{equation}
H=H_{\text{mop}}+\int_{0}^{\infty }dk\omega _{k}c_{k}^{\dag }c_{k}+\xi
\int_{0}^{\infty }dk(a^{\dag }c_{k}+c_{k}^{\dag }a),
\label{hamiltonian}
\end{equation}
where $c_{k}$ ($c_{k}^{\dag }$) is the annihilation (creation) operator of the $k$th mode in the continuous fields, with the resonance frequency $\omega_{k}=c|k|$ ($c$ is the velocity of light in vacuum). The $\xi$ term in Eq.~(\ref{hamiltonian}) represents the photon-hopping coupling between the continuous fields outside the cavity and the optomechanical cavity field, and $\xi $ is the coupling strength. In typical optomechanical systems, the decay rate of the cavity field is much larger than the decay rate of the mechanical mode. Therefore, we neglect the mechanical dissipation in the two-photon scattering problem because the photon scattering can be completed in a short period during which the influence on the photons caused by the mechanical dissipation can be neglected.

\section{Equations of Motion for probability amplitudes}
In this work, we will study the photon scattering by solving the equations of motion for probability amplitudes within the Wigner-Weisskopf framework. In this system, the total photon number operator in the cavity and these continuous fields outside the cavity is defined by $\hat{N}=a^{\dag }a+\int_{0}^{\infty }dkc_{k}^{\dag }c_{k}$, and this photon number operator $\hat{N}$ is a conserved quantity based on the commutative relation $[\hat{N},H]=0$. For studying the two-photon scattering, we therefore derive the equations of motion for probability amplitudes in the two-photon subspaces.

For below convenience, we work in a rotating frame with respect to $H_{0}=\omega _{c}(a^{\dag }a+\int_{0}^{\infty }dkc_{k}^{\dag }c_{k})$. In
this rotating frame, the Hamiltonian $H$ given in Eq.~(\ref{hamiltonian}) becomes
\begin{equation}
H_{I}=H_{\text{mop}}^{I}+\int_{0}^{\infty }dk\Delta _{k}c_{k}^{\dag
}c_{k}+\xi \int_{0}^{\infty }dk(a^{\dag }c_{k}+c_{k}^{\dag }a),
\label{Interhamiltonian}
\end{equation}
where $\Delta _{k}=\omega _{k}-\omega _{c}$ is the detuning and $H_{\text{mop}}^{I}$ takes the form as
\begin{equation}
H_{\text{mop}}^{I}=\omega _{M}b^{\dag }b+g_{1}a^{\dag }a(b^{\dag
}+b)+g_{2}a^{\dag }a(b^{\dag }+b)^{2}.
\end{equation}%
The eigensystem of the Hamiltonian $H_{\text{mop}}^{I}$ is given by $H_{\text{mop}}^{I}\left\vert m\right\rangle _{a}|\tilde{j}\left( m\right)
\rangle _{b}=E_{m,j}^{\prime }\left\vert m\right\rangle _{a}|\tilde{j}\left(m\right) \rangle _{b}$, where the eigenvalues are given by $E_{m,j}^{\prime }=-g_{1}^{2}e^{-4r_{m}}m^{2}/\omega _{M}+j\omega_{M}e^{2r_{m}}+\omega _{M}(e^{2r_{m}}-1)/2$.

In the two-photon subspaces, an arbitrary pure state of the whole system can be expressed as
\begin{eqnarray}
\left\vert \Phi \left( t\right) \right\rangle  &=&\sum_{j=0}^{\infty
}A_{j}\left( t\right) \left\vert 2\right\rangle _{a}\left\vert \emptyset
\right\rangle |\tilde{j}\left( 2\right) \rangle _{b}  \nonumber \\
&&+\sum_{j=0}^{\infty }\int_{0}^{\infty }dkB_{j,k}\left( t\right) \left\vert
1\right\rangle _{a}\left\vert 1_{k}\right\rangle |\tilde{j}\left( 1\right)
\rangle _{b}  \nonumber \\
&&+\sum_{j=0}^{\infty }\int_{0}^{\infty }dp\int_{0}^{p}dqC_{j,p,q}\left(
t\right) \left\vert 0\right\rangle _{a}|1_{p},1_{q}\rangle \left\vert
j\right\rangle _{b},
\label{generalstate}
\end{eqnarray}
where $\left\vert 2\right\rangle _{a}\left\vert \emptyset \right\rangle $ denotes the basis state with two photons in the cavity and no photon in these outside fields, $\left\vert 1\right\rangle _{a}\left\vert 1_{k}\right\rangle $ stands for a single photon inside the cavity and the other photon in the $k$th mode of the continuous-mode fields, and the basis state $\left\vert 0\right\rangle _{a}|1_{p},1_{q}\rangle $ describes the situation in which the cavity field has no photon and two photons are in the $p$th and $q$th modes of the continuous fields. In Eq.~(\ref{generalstate}), $A_{j}\left( t\right) $, $B_{j,k}\left( t\right) $, and $C_{j,p,q}\left( t\right) $ denote the corresponding probability amplitudes.

Based on the Schr\"{o}dinger equation, we can obtain equations of motion for these probability amplitudes as
\begin{subequations}
\label{eqofmotion}
\begin{align}
\dot{A}_{j}(t)=& -iE_{2,j}^{\prime }A_{j}(t)  \notag \\
& -i\sqrt{2}\xi \sum_{s=0}^{\infty }\int_{0}^{\infty }\,_{b}\langle \tilde{j}%
\left( 2\right) \left\vert \tilde{s}(1)\right\rangle _{b}B_{s,k}(t)dk, \\
\dot{B}_{j,k}(t)=& -i\left( E_{1,j}^{\prime }+\Delta _{k}\right)
B_{j,k}\left( t\right)   \notag \\
& -i\sqrt{2}\xi \sum_{s=0}^{\infty }A_{s}\left( t\right) \,_{b}\langle
\tilde{j}\left( 1\right) |\tilde{s}\left( 2\right) \rangle _{b} \notag \\
& -i\xi \sum_{s=0}^{\infty }\,_{b}\langle \tilde{j}\left( 1\right) |s\rangle
_{b}\int_{0}^{\infty }C_{s,p,k}\left( t\right) dp, \\
\dot{C}_{j,p,q}(t)=& -i(j\omega _{M}+\Delta _{q}+\Delta _{p})C_{j,p,q}\left(
t\right)   \notag \\
& -i\xi \sum_{s=0}^{\infty }\,_{b}\langle j|\tilde{s}\left( 1\right) \rangle
_{b}[B_{s,p}(t)+B_{s,q}(t)].
\end{align}
\end{subequations}
According to the above equations, we can see that the generalized Franck-Condon factors $\,_{b}\langle \tilde{j}\left( 2\right) \left\vert \tilde{s}\left(1\right)
\right\rangle_{b}$, $\,_{b}\langle \tilde{j}\left( 1\right) |\tilde{s}\left( 2\right) \rangle_{b}$, $\,_{b}\langle \tilde{j}\left( 1\right)
|s\rangle _{b}$, and$\ \,_{b}\langle j|\tilde{s}\left( 1\right) \rangle _{b}$ determine the transition rates related to the photon scattering
processes. These factors can be calculated based on the relation $_{b}\langle \tilde{j}(m)|\tilde{s}(n)\rangle _{b}=\,_{b}\langle
j|S(r_{n}-r_{m})D\{\alpha _{n}-\alpha _{m}[\cosh (r_{n}-r_{m})+\sinh(r_{n}-r_{m})]\}|s\rangle _{b}$. Here, the inner product between the number states $|s\rangle _{b}$ and the squeezed displaced number states $|\tilde{j}\rangle _{b}$ of the mechanical mode is determined by the relation~\cite{{JMO1990}}
\begin{eqnarray}
&&\,_{b}\langle s|S\left( r\right) D\left( \alpha \right) |j\rangle _{b}
\notag \\
&=&\frac{1}{\left( s!j!\mu \right) ^{\frac{1}{2}}}\left( \frac{\nu }{2\mu }%
\right) ^{\frac{s}{2}}e^{\left( -\frac{\left\vert \alpha \right\vert ^{2}}{2}%
+\frac{\nu ^{\ast }}{2\mu }\alpha ^{2}\right) }  \notag \\
&&\times \sum_{k=0}^{\min \left( j,s\right) }\frac{C_{j}^{k}2^{k}s!}{\left(
s-k\right) !}\left( 2\mu \nu \right) ^{-\frac{k}{2}}H_{s-k}\left( \frac{%
\alpha }{\sqrt{2\mu \nu }}\right)  \notag \\
&&\times \left( -\frac{\nu ^{\ast }}{2\mu }\right) ^{\frac{j-k}{2}%
}H_{j-k}\left( \frac{\alpha \nu ^{\ast }-\alpha ^{\ast }\mu }{\sqrt{-2\mu
\nu ^{\ast }}}\right),
\end{eqnarray}%
where we introduce the variables $\mu =\cosh R$ and $\nu =e^{-i\theta }\sinh R$, with $R$ and $\theta $ defined by $r=Re^{i\theta }$.

\section{Two-Photon Scattering Solution}

In the two-photon scattering case, the cavity field is initially in a vacuum state and the two photons are in the outside fields, while the mechanical mode could be in an arbitrary state. To obtain the analytical solution of the two-photon scattering, we consider the situation where the two photons are initially in a Lorentzian wavepacket. Below we will first solve the equation of motion with the Laplace transform method when the mechanical mode is initially in the number state $|n_{0}\rangle _{b}$. Once the solution in this case is obtained, the solution corresponding to an arbitrary initial state of the mechanical mode can be obtained by superposition.

In the two-photon scattering case, the initial conditions for these probability amplitudes are given by $A_{j}(0)=0$, $B_{j,k}(0)=0$, and
\begin{equation}
C_{j,p,q}\left( 0\right) =\frac{G\delta _{j,n_{0}}}{(\Delta _{p}-\Delta
_{1}+i\epsilon )(\Delta _{q}-\Delta _{2}+i\epsilon )}+\Delta
_{1}\leftrightarrow \Delta _{2},  \label{amplitude}
\end{equation}%
where the normalization constant $G$ is given by%
\begin{equation}
G=\frac{\epsilon }{\pi }\left( 1+\frac{4\epsilon ^{2}}{(\Delta _{1}-\Delta
_{2})^{2}+\left( 2\epsilon \right) ^{2}}\right) ^{-\frac{1}{2}},
\end{equation}%
with $\Delta _{i=1\text{,}2}=\omega _{i}-\omega _{c}$ and $\epsilon $\ being the detuning and spectral width of the two-photon wavepacket, respectively.

The transient solution of these probability amplitudes $A_{n_{0},j}(t)$, $B_{n_{0},j,k}(t)$, and $C_{n_{0},j,p,q}\left( t\right) $ can be obtained with the Laplace transform method. Here, we have added the subscript $n_{0}$ in the transient solution to mark the initial state $|n_{0}\rangle _{b}$ of the mechanical mode. For studying the two-photon scattering, we focus on the long-time solution of the system. After a lengthy calculation, we obtain the long-time solution as $A_{n_{0},j}\left(\infty \right) =0$, $B_{n_{0},j,k}\left( \infty \right) =0$, and%
\begin{eqnarray}
&&C_{n_{0},j,p,q}\left( \infty \right)  \notag \\
&=&G\left[ (C_{1}+C_{2}+C_{3}+C_{4})+(\Delta _{p}\leftrightarrow \Delta _{q})%
\right] e^{-i(\Delta _{p}+\Delta _{q}+j\omega _{M})t},  \notag \\
&&  \label{solution}
\end{eqnarray}%
which indicate that the two photons exit completely out of the cavity in the long-time limit. In Eq.~(\ref{solution}), the amplitude components $C_{1}$, $C_{2}$, $C_{3}$, and $C_{4}$ are given by
\begin{subequations}
\label{C1-4}
\begin{align}
C_{1} =&\frac{1}{\Delta _{p}-\Delta _{1}+i\epsilon }\frac{1}{\Delta
_{q}-\Delta _{2}+i\epsilon }\delta _{j,n_{0}},  \\
C_{2} =&\sum_{s=0}^{\infty }\frac{-i\gamma _{c}F_{2}}{M_{1}M_{2}(\Delta
_{p}-\Delta _{2}+i\epsilon )}+\Delta _{1}\leftrightarrow \Delta _{2}, \\
C_{3} =&\sum_{s,s^{\prime },l=0}^{\infty }\frac{-\gamma _{c}^{2}F_{3}}{%
M_{1}M_{3}M_{4}M_{5}}+\Delta _{1}\leftrightarrow \Delta _{2},  \\
C_{4} =&\sum_{s,s^{\prime },l=0}^{\infty }\frac{-2\gamma _{c}^{2}F_{4}}{%
M_{1}M_{3}M_{4}M_{6}}+\Delta _{1}\leftrightarrow \Delta _{2},
\end{align}
\end{subequations}
which correspond to four different physical processes. In Eq.~(\ref{C1-4}), we introduce the transition chain coefficients
\begin{subequations}
\begin{align}
F_{2} =&\,_{b}\langle j|\tilde{s}\left( 1\right) \rangle _{bb}\langle
\tilde{s}\left( 1\right) |n_{0}\rangle _{b},  \\
F_{3} =&\,_{b}\langle j|\tilde{s}\left( 1\right) \rangle _{bb}\langle
\tilde{s}\left( 1\right) |s^{\prime }\rangle _{bb}\langle s^{\prime }|\tilde{%
l}\left( 1\right) \rangle _{bb}\langle \tilde{l}\left( 1\right)
|n_{0}\rangle _{b},  \\
F_{4} =&\,_{b}\langle j|\tilde{s}\left( 1\right) \rangle _{bb}\langle
\tilde{s}\left( 1\right) |\tilde{s}^{\prime }\left( 2\right) \rangle
_{bb}\langle \tilde{s}^{\prime }\left( 2\right) |\tilde{l}\left( 1\right)
\rangle _{bb}\langle \tilde{l}\left( 1\right) |n_{0}\rangle _{b},  \notag \\
\label{transition}
\end{align}
\end{subequations}
which denote the transition processes of the mixed cavity optomechanical system. The resonance conditions related to the photon absorptions and emissions are determined by the poles of these denominators in Eq.~(\ref{C1-4}). Namely, these resonance conditions are governed by the relations Re[$M_{1-6}$]=0, where the expressions of the variables $M_{1-6}$ are given by
\begin{subequations}
\begin{align}
M_{1} =&\Delta _{q}-E_{1,s}^{\prime }+j\omega _{M}+i\frac{\gamma _{c}}{2},
 \\
M_{2} =&\Delta _{q}-\Delta _{1}+(j-n_{0})\omega _{M}+i\epsilon ,   \\
M_{3} =&\Delta _{p}+\Delta _{q}-\Delta _{1}-\Delta _{2}+\left(
j-n_{0}\right) \omega _{M}+2i\epsilon ,   \\
M_{4} =&\Delta _{p}+\Delta _{q}-\Delta _{1}-E_{1,l}^{\prime }+j\omega
_{M}+i(\epsilon +\frac{\gamma _{c}}{2}), \\
M_{5} =&\Delta _{q}-\Delta _{1}+(j-s^{\prime })\omega _{M}+i\epsilon ,
 \\
M_{6} =&\Delta _{p}+\Delta _{q}+j\omega _{M}-E_{2,s^{\prime }}^{\prime
}+i\gamma _{c},  \label{resonance}
\end{align}
\end{subequations}
with $\gamma _{c}=2\pi \xi ^{2}$ being the cavity-field decay rate.

To understand the two-photon scattering processes, below we analyze the four amplitudes $C_{1-4}$ given in Eq.~(\ref{solution}). The first term $C_{1}$ shows the process in which the two photons do not enter the cavity. In this case, the two photons are reflected directly by the coupling ending mirror, which provides the photon-hopping channel.

The second term $C_{2}$ shows the transition process $|0\rangle_{a}|n_{0}\rangle _{b}\rightarrow |1\rangle _{a}|\tilde{s}(1)\rangle_{b}\rightarrow |0\rangle _{a}|j\rangle _{b}$, which corresponds to the resonance conditions $n_{0}\omega _{M}+\omega _{1}=E_{1,s}$ and $\omega_{q}=E_{1,s}-j\omega _{M}$ [Re$(M_{1}-M_{2})=0$ and Re$(M_{1})=0$]. This means that only one photon enters the cavity, and the other photon is reflected by the coupling ending mirror. This phenomenon represents a single-photon scattering and reflection process. The frequency of the emitted photon is governed by the resonance condition
\begin{equation}
\Delta _{q}=s\omega _{M}e^{2r_{1}}-j\omega _{M}-\delta,
\end{equation}%
with%
\begin{equation}
\delta =\frac{g_{1}^{2}e^{-4r_{1}}}{\omega _{M}}-\frac{\omega _{M}}{2}%
(e^{2r_{1}}-1)
\end{equation}%
being the ground-state energy shift of the resonator induced by the single-photon squeezing and displacement (see Fig.~\ref{energylevel}).

The third term $C_{3}$ indicates that the mixed optomechanical system experiences the following transitions: $|0\rangle _{a}|n_{0}\rangle _{b}\rightarrow|1\rangle _{a}|\tilde{l}(1)\rangle _{b}\rightarrow |0\rangle _{a}|s^{\prime}\rangle _{b}\rightarrow |1\rangle _{a}|\tilde{s}(1)\rangle _{b}\rightarrow|0\rangle _{a}|j\rangle _{b}$. The photon resonance conditions are determined by there relations $\omega_{1}+s^{\prime }\omega _{M}=E_{1,s}$ [i.e., $\text{Re}(M_{1}-M_{5})=0$], $n_{0}\omega _{M}+\omega _{2}=E_{1,l}$ [$\text{Re}(M_{4}-M_{3})=0$], $\Delta _{p}=l\omega_{M}e^{2r_{1}}-s^{\prime }\omega _{M}-\delta $ [$\text{Re}(M_{4}-M_{5})=0$], and $\Delta _{q}=s\omega _{M}e^{2r_{1}}-j\omega _{M}-\delta $ [$\text{Re}(M_{1})=0$]. It describes the process in which the first photon is emitted out of the cavity, then the second photon enters the cavity. It can be seen from $F_{3}$ that the maximum photon number in the cavity is $1$. Therefore, $C_{3}$ describes a successive single-photon scattering process.

The fourth term $C_{4}$ corresponds to the transitions: $|0\rangle_{a}|n_{0}\rangle _{b}\rightarrow |1\rangle _{a}|\tilde{l}(1)\rangle_{b}\rightarrow |2\rangle _{a}|\tilde{s}^{\prime }(2)\rangle _{b}\rightarrow|1\rangle _{a}|\tilde{s}(1)\rangle _{b}\rightarrow |0\rangle _{a}|j\rangle_{b}$. The resonance conditions for this process are $n_{0}\omega_{M}+\omega _{2}=E_{1,l}$ [$\text{Re}(M_{4}-M_{3})=0$], $\omega_{1}+E_{1,l}=E_{2,s^{\prime }}$ [$\text{Re}(M_{6}-M_{4})=0$], $%
\omega _{p}=E_{2,s^{\prime }}-E_{1,s}$ [Re$(M_{6}-M_{1})=0$], and $\omega_{q}=E_{1,s}-j\omega _{M}$ [$\text{Re}(M_{1})=0$]. This transition chain indicates that the two photons exist out of the cavity at the same time, which describes a `genuine' two-photon scattering process. The frequencies of the emitted photons are governed by the resonance conditions
\begin{subequations}
\begin{eqnarray}
\left. \Delta _{p}=s^{\prime }\omega _{M}e^{2r_{2}}-s\omega
_{M}e^{2r_{1}}+\delta -\nu ,\right.  \\
\left. \Delta _{p}+\Delta _{q}=s^{\prime }\omega _{M}e^{2r_{2}}-j\omega
_{M}-\nu ,\right.
\end{eqnarray}
\end{subequations}
with%
\begin{equation}
\nu =\frac{4g_{1}^{2}e^{-4r_{2}}}{\omega _{M}}-\frac{\omega _{M}}{2}%
(e^{2r_{2}}-1)
\end{equation}%
being the ground-state energy shift of the resonator induced by two-photon squeezing and displacement (see Fig.~\ref{energylevel}).
\begin{figure}[tbp]
\center
\includegraphics[width=0.48 \textwidth]{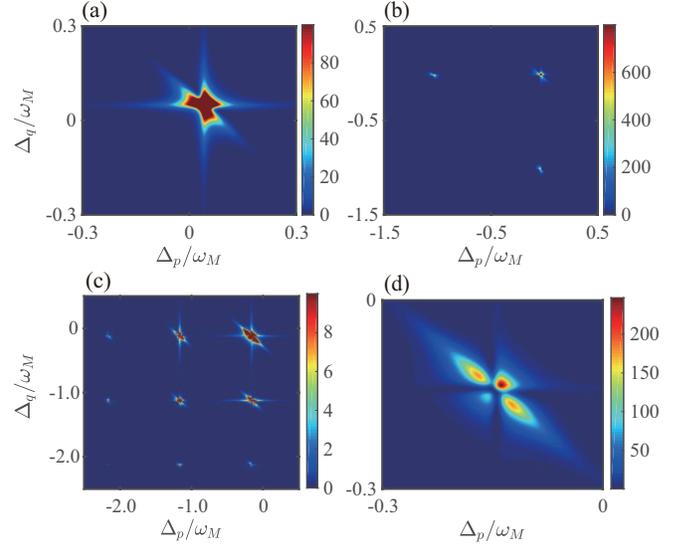}
\caption{(Color online) Two-photon scattering spectrum $S(\Delta _{p},\Delta_{q})$ as a function of the photon frequency detunings $\Delta _{p}/\protect\omega _{M}$ and $\Delta _{q}/\protect\omega _{M}$ for the initial mechanical ground state $\left\vert 0\right\rangle_{b}$ when the coupling strengths take various values: (a) $g_{1}/\protect\omega _{M}=0.2$ and $g_{2}/\protect\omega _{M}=0.08$, (b) $g_{1}/\protect\omega _{M}=0.2$ and $g_{2}/\protect\omega _{M}=0.01$, and (c) $g_{1}/\protect\omega _{M}=0.4$ and $g_{2}/\protect\omega _{M}=0.01$. (d) The zoomed view of the peak with the center position $\Delta_{p}+\Delta_{q}=-\delta$ in panel (c). Other parameters are $\protect\gamma _{c}/\protect\omega _{M}=0.1$, $\protect\epsilon /\protect\omega _{M}=0.01$, and $\Delta _{1}=\Delta _{2}=-\delta $.}
\label{Fig2}
\end{figure}
\begin{figure*}[tbp]
\center
\includegraphics[width=0.98 \textwidth]{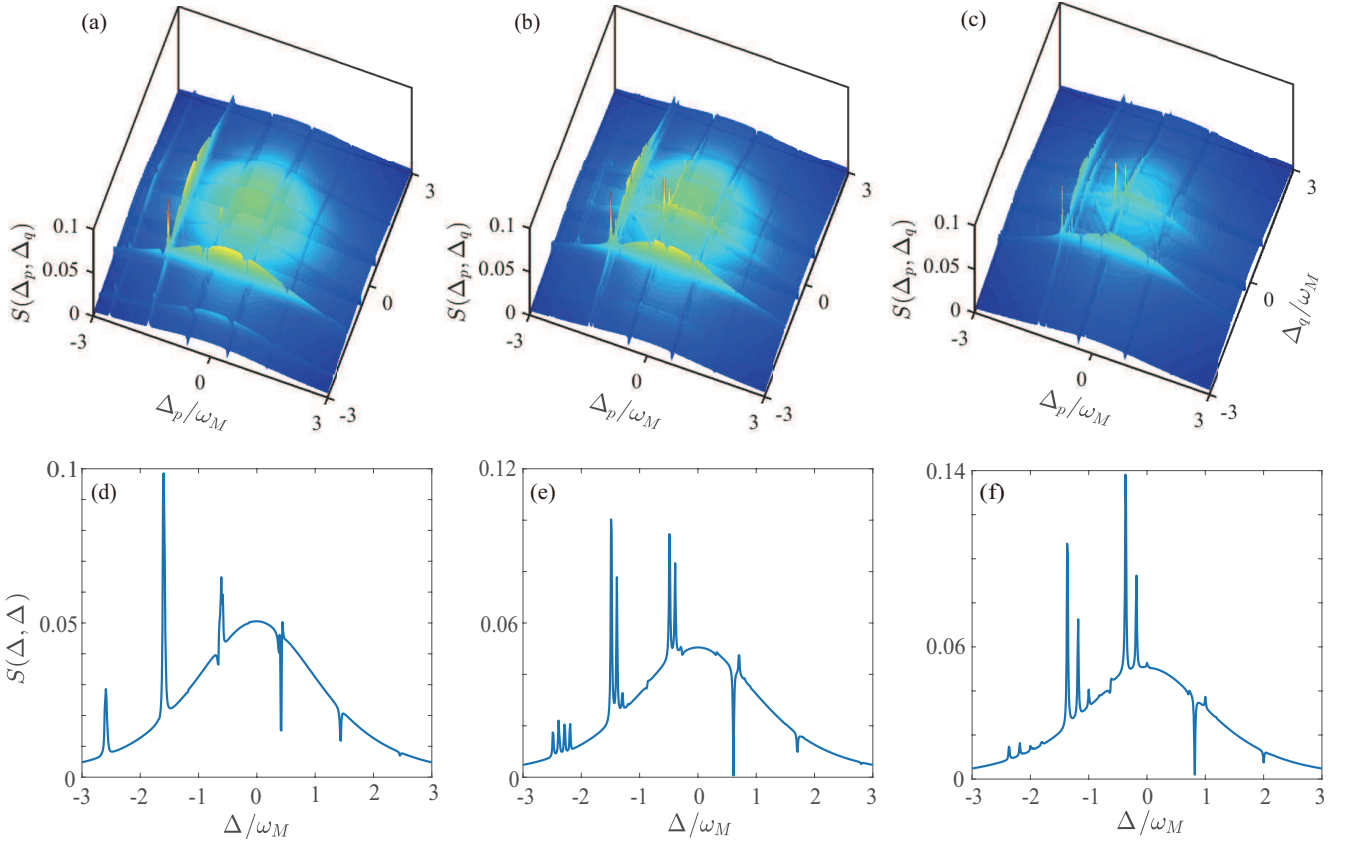}
\caption{(Color online) Two-photon scattering spectrum $S(\Delta_{p},\Delta_{q})$ as a function of $\Delta _{p}/\protect\omega _{M}$ and $\Delta _{q}/\protect\omega _{M}$ for various values of the coupling strength $g_{2}$, when the mechanical mode is in the ground state $\left\vert 0\right\rangle_{b}$. Here, (a), (b), and (c) correspond to $g_{2}/\protect\omega _{M}=0.01$, $0.05$, and $0.10$, respectively. (d-f) The joint spectrum $S(\Delta,\Delta)$ along the diagonal line $\Delta _{p}=\Delta _{q}=\Delta$ in the above corresponding panels. Other parameters are given by $g_{1}/\protect\omega _{M}=0.8$, $\protect\gamma _{c}/\protect\omega_{M}=0.02 $, and $\protect\epsilon /\protect\omega _{M}=2.0$.}
\label{Scatteringspectrum}
\end{figure*}
\section{TWO-PHOTON SCATTERING SPECTRA}
To characterize the photon correlation induced by the two-photon scattering process, we investigate the two-photon scattering spectrum. Quantum correlation between the two scattered photons can be observed by analyzing the shape of the scattering spectrum in the frequency space. Mathematically, quantum correlation between the two scattered photons can be confirmed based on the fact that the probability amplitude $C_{n_{0},j,p,q}$ in Eq.~(\ref{solution}) is not a factorizable function of $\Delta _{p}$ and $\Delta _{q}$. In the long-time limit, the state of the whole system corresponding to the initial state $\left\vert n_{0}\right\rangle _{b}$ of the mechanical mode reads
\begin{equation}
\left\vert \Phi _{n_{0}}\left( \infty \right) \right\rangle
=\sum_{j=0}^{\infty }\int_{0}^{\infty }dp\int_{0}^{p}dqC_{n_{0},j,p,q}\left(
\infty \right) \left\vert 0\right\rangle _{a}|1_{p},1_{q}\rangle \left\vert
j\right\rangle _{b}.
\end{equation}
When the mechanical resonator is initially in a pure state $\left\vert \varphi \right\rangle_{b}=\sum_{n_{0}=0}^{\infty}c_{n_{0}}\left\vert n_{0}\right\rangle _{b}$ or a mixed state $\rho _{b}=\sum_{n_{0}=0}^{\infty }P_{n_{0}}|n_{0}\rangle _{bb}\langle n_{0}|$, the corresponding two-photon scattering spectrum~\cite{Liao2013} can be expressed, respectively, as
\begin{subequations}
\begin{align}
S(\Delta _{p},\Delta _{q})=& \sum_{j=0}^{\infty }\left\vert
\sum_{n_{0}=0}^{\infty }c_{n_{0}}C_{n_{0},j,p,q}\left( \infty \right)
\right\vert ^{2}, \\
S(\Delta _{p},\Delta _{q})=& \sum_{j=0}^{\infty }\sum_{n_{0}=0}^{\infty
}P_{n_{0}}\left\vert C_{n_{0},j,p,q}\left( \infty \right) \right\vert ^{2}.
\end{align}
\end{subequations}
\begin{figure}[t!]
\center
\includegraphics[width=0.45 \textwidth]{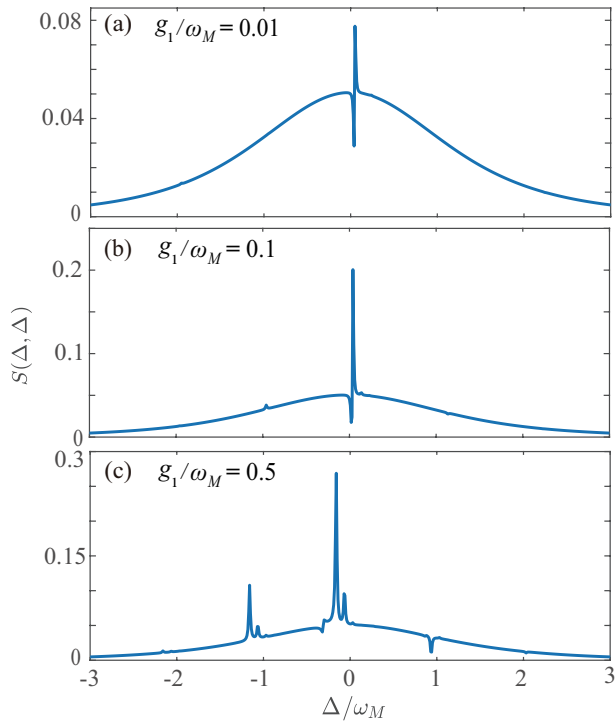}
\caption{(Color online) The two-photon scattering spectrum $S(\Delta,\Delta)$ along the diagonal line $\Delta _{p}=\Delta _{q}=\Delta$ when the coupling strength $g_{1}$ takes various values: (a) $g_{1}/\protect\omega _{M}=0.01$, (b) $g_{1}/\protect\omega _{M}=0.1$, and (c) $g_{1}/\protect\omega _{M}=0.5$. Other parameters are given by $g_{2}/\protect\omega _{M}=0.05$, $\protect\gamma _{c}/\protect\omega_{M}=0.02 $, and $\protect\epsilon /\protect\omega _{M}=2.0$.}
\label{g1}
\end{figure}
\begin{figure}[t!]
\center
\includegraphics[width=0.45 \textwidth]{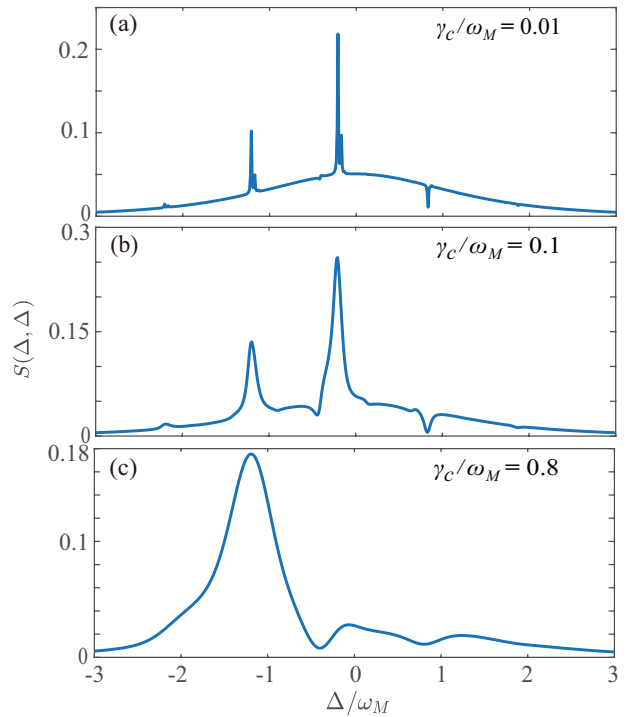}
\caption{(Color online) The two-photon scattering spectrum $S(\Delta,\Delta)$ along the diagonal line $\Delta _{p}=\Delta _{q}=\Delta$ at various values of $\gamma_{c}$: (a) $\protect\gamma _{c}/\protect\omega_{M}=0.01$, (b) $\protect\gamma _{c}/\protect\omega_{M}=0.1$, and (c) $\protect\gamma _{c}/\protect\omega_{M}=0.8$. Other parameters are $g_{1}/\protect\omega _{M}=0.5$, $g_{2}/\protect\omega _{M}=0.02$, and $\protect\epsilon /\protect\omega _{M}=2.0$.}
\label{Gammac}
\end{figure}

Below, we analyze the two-photon scattering spectrum in various cases. In Fig.~\ref{Fig2}, we plot the two-photon scattering spectrum as a function of the photon frequency detunings $\Delta _{p}$ and $\Delta _{q}$ when the initial state of the mechanical mode is $\left\vert 0\right\rangle _{b}$. To exhibit the phonon sideband effect in this system, we first consider the weak-coupling case in which the probability amplitude $C_{n_{0},j,p,q}(\infty )$ can be expanded upto the lower orders of $r_{m}$ and $\alpha _{m}$. Meanwhile, we assume that the system works in the single-photon strong-coupling regime $g_{1}>\gamma_{c}$ and the resolved-sideband regime $\omega _{M}>\gamma _{c}$, such that the phonon sideband peaks can be resolved in the scattering spectrum. In Fig.~\ref{Fig2}(a), we can see that the spectrum only shows one main scattering peak, and there is no phonon sidebands. With the increase of the parameters $r_{m}$ and $\alpha _{m}$, more and more phonon sideband peaks appear in the spectrum, as shown in Figs.~\ref{Fig2}(b) and \ref{Fig2}(c). In particular, we show the zoomed view of the main peak [centered at $\Delta_{p}+\Delta_{q}=-\delta$ in Fig.~\ref{Fig2}(c)] and found that the two photons are frequency anticorrelated with a probability concentrated along the line parallel to $\Delta_{p}+\Delta_{q}=0$, as shown in Fig.~\ref{Fig2}(d).

In the two-photon scattering case, the width of the injected two-photon Lorentzian wavepacket is a controllable variable. Next we will focus on the case of wide wavepacket injection, i.e., $\epsilon /\omega _{M}>1$.  In Fig.~\ref{Scatteringspectrum}, we plot the scattering spectrum as a function of $\Delta _{p}/\protect\omega _{M}$ and $\Delta _{q}/\protect\omega _{M}$ in the wide wavepacket case when the parameter $g_{2}$ takes various values. In the mixed cavity optomechanical system, the two optomechanical coupling strengths $g_ {1}$ and $g_ {2}$ take important roles in the spectrum features. It has been found that, in the single-photon scattering case, the single-photon strong-coupling condition $g_ {1}>\gamma _{c}$ is a necessary parameter condition for the exhibition of these phonon sideband peaks (the distance between neighboring phonon sideband peaks is $\omega_{M}$)~\cite{Zhou2019}. In addition, when $g_{2}/\protect\omega _{M}\ll 1$, the parameter condition $2g_ {2}>\gamma _{c}$ is the necessary condition for observation of subpeaks in each phonon sideband. Note that the subpeaks could be either peaks or dips caused by quantum interference effect. Based on the above analyses, in Fig.~\ref{Scatteringspectrum} we choose different values of $g_{2}$ to show the gradual transition from the subpeak-unresolved regime to the subpeak-resolved regime. Concretely, in Figs.~\ref{Scatteringspectrum}(a-c) we plot the scattering spectrum as a function of $\Delta _{p}$ and $\Delta _{q}$. Here, we can see that there are many sidebands in the spectrum (the grid in the two dimension diagram). In particular, we can see many subpeaks in the spectrum when the parameter condition $2g_ {2}>\gamma _{c}$ is satisfied.

To clearly see the phonon sideband peaks and subpeaks in the spectrum, in Figs.~\ref{Scatteringspectrum}(d-f) we plot the two-photon scattering spectrum along the diagonal line, i.e., we take $\Delta _{p}=\Delta _{q}=\Delta$ in the two-photon spectrum. Here, the phonon sideband peaks can be seen because the condition $g_{1}>\gamma _{c}$ is satisfied in panels \ref{Scatteringspectrum}(d-f). In addition, the subpeaks can be observed in panels \ref{Scatteringspectrum}(e-f) because the condition $2g_ {2}>\gamma _{c}$ is satisfied only in these two cases. In particular, we can see both peaks and dips in the spectrum. This is because quantum interference exists between the two-photon reflection channel and the scattering channel.

To further analyze the parameter condition for resolving these phonon sideband peaks, in Fig.~\ref{g1} we plot the two-photon scattering spectrum along the diagonal line $\Delta _{p}=\Delta _{q}=\Delta$ at various values of $g_{1}/\protect\omega _{M}$. Here, we choose $2g_ {2}>\gamma _{c}$ and hence the subpeaks can be observed in each sideband. For the parameter $g_{1}$, we find that the phonon sidebands can be observed only when the condition $g_ {1}>\gamma _{c}$ is satisfied. In Fig.~\ref{g1}, the phonon sidebands can only be observed in panels (b) and (c). In addition, for a large value of $g_ {1}$, more phonon sideband peaks can be observed in the spectrum, and the subpeaks can also be observed more clearly.

We also study the spectral features when the cavity-field decay rate takes various values. As shown in Fig.~\ref{Gammac}, we can see both the phonon sideband peaks and the subpeaks in panel (a) ($g_ {1}>\gamma _{c}$, $2g_ {2}>\gamma _{c}$). In panel (b), the spectrum shows phonon sideband peaks, without the subpeaks ($g_ {1}>\gamma _{c}$, $2g_ {2}<\gamma _{c}$). In panel (c), both the phonon sideband peaks and the subpeaks cannot be observed, this is because the system works in the parameter regime of $g_ {1}<\gamma _{c}$ and $2g_ {2}<\gamma _{c}$.

\section{CONCLUSION}
In conclusion, we have studied two-photon scattering in the mixed cavity optomechanical system. With the Laplace transform method, we have obtained the exact analytical solution of two-photon scattering in the long-time limit, and then the two-photon scattering spectrum can be obtained. We have found that the two scattered photons are frequency anticorrelated. In particular, we have analyzed the two-photon scattering spectrum along the diagonal line, and found the relationship between the system parameters and the spectral features. The results indicated that the conditions for resolution of the phonon sideband peaks and the subpeaks around each phonon sideband are $g_ {1}>\gamma _{c}$ and $2g_ {2}>\gamma _{c}$ ($g_{2}/\protect\omega _{M}\ll 1$), respectively. Therefore, this work not only provides a scattering method to create correlated photon pairs, but also presents a spectrometric method to characterize the parameters of the mixed cavity optomechanical system.

\begin{acknowledgments}
J.-F.H. is supported in part by the National Natural Science Foundation of China (Grant No. 12075083), Scientific Research Fund of Hunan Provincial Education Department (Grant No. 18A007), and Natural Science Foundation of Hunan Province, China (Grant No. 2020JJ5345). J.-Q.L. is supported in part by National Natural Science Foundation of China (Grants No. 11774087, No. 11822501, and No. 11935006), Hunan Science and Technology Plan Project (Grant No. 2017XK2018), and the Science and Technology Innovation Program of Human of Hunan Province (Grant No. 2020RC4047).
\end{acknowledgments}

\end{document}